\documentstyle[preprint,amsfonts,aps,prd,epsf,floats]{revtex}

\newcommand{\beq}{\begin{equation}}
\newcommand{\eeq}{\end{equation}}
\newcommand{\beqa}{\begin{eqnarray}}
\newcommand{\eeqa}{\end{eqnarray}}
\newcommand{\non}{\nonumber}

\begin{document}

\draft

\title{
\hfill {\rm RU99-12-B} \\
\hfill {\rm OUTP-99-65-P} \\
Gauge Invariant Variational Approach with Fermions: the Schwinger Model
}

\author{
William E. Brown\thanks{E-mail: wbrown@theory.rockefeller.edu}
}

\address{
Theoretical Physics, The Rockefeller
University, 1230 York Avenue - New York, NY 10021
}

\author{
Juan P. Garrahan\thanks{E-mail: j.garrahan1@physics.ox.ac.uk},
Ian I. Kogan\thanks{E-mail: i.kogan1@physics.ox.ac.uk} 
and Alex Kovner\thanks{E-mail: a.kovner1@physics.ox.ac.uk}
}

\address{
Theoretical Physics, University of Oxford,
1 Keble Road, Oxford, OX1 3NP, UK.
}

\date{DRAFT VERSION: \today}

\maketitle

\begin{abstract}
We extend the gauge invariant variational approach of \cite{KK95} to
theories with fermions. As the simplest example we consider the
massless Schwinger model in 1+1 dimensions. We show that in this
solvable model the simple variational calculation gives exact results.

\end{abstract}


\section{Introduction.}

This paper is an additional step in our exploration of 
the gauge invariant
variational approach suggested in \cite{KK95}.
This approach was developed as an attempt to study analytically the
nonperturbative infrared dynamics of nonabelian gauge theory. The idea
is that keeping gauge invariance exact at all stages of the
calculation may be crucial for consistent understanding of the ground
state structure of QCD.
One therefore tries to account for the generation of nonperturbative QCD
condensates by introducing them as variational parameters in
the explicitly gauge invariant (Gaussian like) trial wave functional.

So far this variational approach has been applied to pure gluodynamics.
There it seems to capture many of the essential
features; mass generation, formation of the gluon
condensate, \cite{KK95}, and asymptotic freedom,
\cite{BK97,B97,diakonov}.
Instanton transitions have been analyzed in
\cite{brogakoko} and are identified with the saddle points in the
integration over the gauge group which projects the Gaussian wave
functional onto the gauge invariant physical Hilbert space.
Interestingly we found that in the best variational state 
instantons of large size 
are suppressed and the large size instanton
problem that arises in the standard WKB calculation is completely
avoided. The dynamically
generated mass present in the best variational state stabilizes
instantons at a size $\rho \sim 1/M$.

Although there remain many open questions regarding the reliability
of these
calculations, it is interesting to see how a simple Ansatz of the same
type would do in a theory with fermions. In particular one would like
to see whether including fermions leads within the gauge invariant
variational framework to the generation of fermionic condensates and
chiral symmetry breaking.

As a first step
towards this goal, in this paper we will study the simple toy model,
1+1 dimensional Abelian theory with massless fermions, the so-called
Schwinger model \cite{S62}. Our aim is not  to further the understanding
of the model itself (which has been solved exactly  many times by 
different methods and is well understood) 
but to develop a workable generalization of our
method to fermionic theory and test it in the simplest
possible but nontrivial setting.

One of the most striking similarities between the Schwinger model and
QCD is that in neither theory are the asymptotic states represented by
the fields of the Lagrangian.  The Schwinger model, with massless
fermions, exhibits complete screening of charges and has one neutral
asymptotic state; a `meson' of mass $e/\sqrt{\pi}$.  
The presence and interconnectedness of the axial anomaly, instantons
and the massive asymptotic states of the theory has made the Schwinger
model a rich field of study.  

The outline of this paper is as follows. In Section 2 we generalize
the gauge invariant Gaussian variational Ansatz to include fermions
and perform the variational calculation. We find that the best
variational state in the
fermionic sector does not contain 
new nontrivial
structure. 
In Section 3 we cross check our
calculation using the bosonized form of the model. We find that
our best trial state is in fact the exact ground state of the theory.
In Section 4 we conclude by calculating the
fermionic condensate. 
We recover the correct
value of the condensate
and show thereby that the
chiral anomaly is correctly represented in our calculation.

\section{Variational Calculation in the Schwinger Model.}

The Lagrangian density for the Schwinger model reads 
\beq {\cal L} =
\frac{1}{4} F_{\mu \nu} F^{\mu \nu} + \bar{\psi} \, \gamma^{\mu} \left(
i \partial_{\mu} - e A_{\mu} \right) \psi.  
\eeq 
The theory is super-renormalizable and the coupling constant, $e$, has
mass dimension $+1$.  The $2 \times 2$ Dirac matrices obey the usual
anti-commutation relation, $\{\gamma^\mu,\gamma^\nu\}=2g^{\mu\nu}$,
where spatial indices run from $0$ to $1$ and $g^{00}=1$
We write the Schwinger model in the Hamiltonian formalism in the
temporal gauge, $A_0=0$. This leaves one degree of freedom for the
photon field and all spatial indices are now suppressed; 
$A_1(x) = A(x)$ and $\partial_1 = \partial$.  The Hamiltonian is,
\beq H = \int dx \, \left( 
	\frac{1}{2}E^2(x) +
	\bar{\psi}(x) (i\partial - e A(x) )\gamma_1 \psi(x)
	\right) ,
\eeq
with the usual canonical relations,
\beqa
[A(x),E(y)] 
	&=& 
	i \delta(x-y), ~~~~~~ E(x) = \dot{A}(x), 
	\\ \nonumber  
[\psi(x),\pi_{\psi}(y)] 
	&=& 
	i \delta(x-y), ~~~~~~ \pi_{\psi}(x) = i \psi^{\dagger}(x).
\eeqa
The gauge invariance of a physical state, in this case
the vacuum wavefunctional, is ensured by requiring that it satisfies
Gauss' law,
\beq 
{\cal G}(x)\Psi[A,\psi] = 
	\left[\partial E(x) - e \psi^{\dagger}(x)
	\psi(x)\right]\Psi[A,\psi] = 0 .
\eeq
As an Ansatz for the vacuum wave functional of the theory we take
the product
of the vacuum wavefunctionals in the gauge and fermion sectors,
projected onto the gauge invariant subspace.
Following \cite {KK95} for the functional in the gauge sector we take 
a Gaussian with an arbitrary width $G(p)$ which is to be treated 
as a functional variational parameter.  

Next we have to address the Ansatz in the fermion sector.
Since we may expect that the effects of chiral symmetry breaking show
up in the fermion sector as dynamical generation of mass,
we propose as an Ansatz for the ground state a Dirac vacuum of free
massive fermions, which we denote by $|m\rangle$. The mass, or rather
the mass 
function $m(p)$, can
depend on momentum and is to be treated as an additional variational
function. In the absence of the gauge fields the fermionic operators
that
annihilate this state are related to the original fermionic operators
by the well
known Bogolyubov transformation. The state itself is somewhat
similar to the BCS state. The important difference is that the mixing
is not between particles and antiparticles, but rather between left-
and right- moving particles, so that the possible condensates in this
state preserve the charge symmetry but break the chiral symmetry
instead.

The product of the gauge and fermionic vacua is
\beq
\Psi[A,\psi] = \exp \left( - \frac{1}{2} \int dx \, dy \,  
 	A(x) G^{-1}(x-y) A(y) \right) | m \rangle \ .
\label{aa}
\eeq 
The vacuum wavefunctional in the gauge theory must of course be gauge
invariant.  This is achieved by projecting (\ref{aa}) onto the
gauge invariant Hilbert space.  To do this note that 
(\ref{aa}) transforms under the gauge transformation generated by Gauss'
law as
\beqa
\Psi^{\phi}[A,\psi] &=& 
	\exp\left( i\int dx \, \phi(x) {\cal G}(x) 
	\right) \, \Psi[A,\psi] 
	\non \\
	&=&  
	\exp \left( - \frac{1}{2} 
	\int dx \, dy \,  A^{\phi}(x) G^{-1}(x-y) A^{\phi}(y) \right)
	\exp \left( 
	-ie\int dx \, \phi(x) \psi^{\dagger}(x) \psi(x)
	\right)
	| m \rangle \ .
	\non 
\eeqa
where,
\beq 
\label{Aphi}
A^{\phi}(x) = A(x) - \partial \phi(x) \ , 
\eeq
and $\phi$ is the gauge function.  The gauge invariant
Ansatz for the vacuum wavefunctional is therefore written by
integrating $\Psi^{\phi}$ over all possible gauge transformations,
\beqa
\Psi[A,\psi] &=& \int D\phi \,
	\exp \left( - \frac{1}{2} \int dx \, dy \, 
	A^{\phi}(x) G^{-1}(x-y) A^{\phi}(y) \right) 
	\non \\
	&&
	\times
	\exp \left( 
	-ie\int dx \, \phi(x) \psi^{\dagger}(x) \psi(x)
	\right)
	| m \rangle \ .
\label{wfal}
\eeqa

In this formalism, one calculates expectation values of local
operators with the Ansatz for the ground state.  Since we are
interested in the calculation of physical quantities the operators we
shall consider are gauge invariant.  The expectation of
any such operator $O$ can be written as
\beqa
\langle {\mathcal O}(A,\bar{\psi},\psi) \rangle &=& 
	\frac{1}{Z} \int D\phi \, DA \, \langle m| \, 
	\exp \left( - \frac{1}{2} 
	\int dx \, dy \, A(x) G^{-1}(x-y) A(y) \right) 
	\non \\
	& & \times \, 
	{\mathcal O}(A,\bar{\psi},\psi) \,
	\exp \left( - \frac{1}{2}
	\int dx \, dy \,  A^{\phi}(x) G^{-1}(x-y) A^{\phi}(y) \right)
	\non \\
	&&
	\times
	\exp \left(
	-ie\int dx \, \phi(x) \psi^{\dagger}(x) \psi(x)
	\right)
	|m\rangle \ .
\eeqa

First let us consider the norm of the state, $O=1$.  
We first integrate over $A$,
\beqa
Z &=& Z_A \int D\phi \,
	\exp \left( - \frac{1}{4} \int dx \, dy \, 
	\partial\phi(x) G^{-1}(x-y) \partial\phi(y) 
	\right) 
	\non \\
	&& \times 
	\langle m | 
	\exp \left( 
	-ie\int dx \, \phi(x) \psi^{\dagger}(x) \psi(x)
	\right)	
	| m \rangle \ , 
	\\
Z_A &=& {\rm Det}^{1/2}(2 \, G^{-1}) \ .
\eeqa
The integration over the fermions is a little less
straightforward. Remembering that $|m\rangle$ is the vacuum of a
free fermionic Hamiltonian with mass $m$ one can use the standard
trick
to express the fermionic matrix element 
as a path integral. The gauge transformation operator in such
a representation
then has to be understood as an insertion at fixed ``time''.  This 
insertion generates an interaction between the fermions and the gauge 
parameter,
\beqa
&& \langle m | 
	\exp \left( 
	-ie\int dx \, \phi(x) \psi^{\dagger}(x) \psi(x)
	\right)	
	| m \rangle 
	=
	\non \\
	&&
	\int D\bar{\psi} \, D\psi \,
	\exp \left[ i \, S_f
	- i e \int dt \, dx \, \delta(t) \phi(x) 
	\bar{\psi}(x) \gamma_0 \psi(x) \right] ,
	\label{pirulo}
\eeqa
where $S_f$ is the action for free fermions with momentum
dependent mass,
\beq
S_f = 
	\int dt \, dx \, dy \, 
	\bar{\psi}(x,t)	
	\left[
	i \delta(x-y) \gamma^{\mu} \partial_{\mu} - m(x-y)
	\right] 
	\psi(y,t) \ .
\eeq
\begin{figure}[t]
\epsfxsize 5cm
\centerline{\epsffile{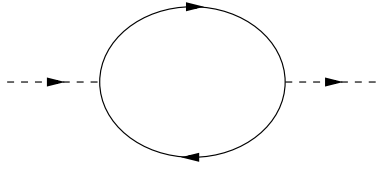}}
\bigskip
\caption{Shows the quadratic term resulting from the expansion in small 
$e \, \phi$.  Solid lines represent a fermion loop and the dashed lines represent the gauge parameter.  Associated with each vertex is the factor $-i e \, 
 \gamma^0 \delta(t)$.}
\label{fig1}
\end{figure}

In the subsequent calculation we will treat $e \, \phi$ as small and
expand to second order the logarithm in the exponential
of the norm produced by integrating over $\psi$ and $\bar{\psi}$.
This results in a series of diagrams.  The leading order is
independent of $\phi$ and contributes a constant factor to the norm which 
will cancel.  Represented by Fig. 1, the term
quadratic in $\phi$ contains a fermion loop and is written,
\beq
-\frac{e^2}{2} \int dx\, dy \,  \phi(x) \phi(y) {\rm Tr} 
\left[
\gamma^0 D_+(x-y) \gamma^0 D_-(y-x) \right],
\eeq
where the equal time propagators are,
\beq
D_{\pm}(k) = - \frac{1}{2} \left[ \gamma^0 \pm \frac{k \gamma^1 + m(k)}{E(k)}
\right],
\eeq
and $E^2(k) = k^2 + m^2(k)$.  We will see later that 
keeping only second 
order terms in $e \, \phi$ is sufficient to locate the best variational
state.
The contribution to the norm from 
fermion integration is,
\beqa
\lefteqn{
\langle m | 
	\exp \left( 
	-ie\int dx \, \phi(x) \psi^{\dagger}(x) \psi(x)
	\right)	
	| m \rangle =} &&
	\non \\
	&& {\rm const.} \times
	\exp\left( - 
	\frac{e^2}{4 \pi} \int dx \, dy \, 
	\partial\phi(x) \, X(x-y) \, \partial\phi(y) 
	+ O\left( (e \phi)^3 \right)\right) \ ,
\label{pepi}
\eeqa
where,
\beq
X(k) = k^{-2} \int \frac{dp}{2 \pi} 
	\left( 1 - \frac{p(p-k) + m(p) m(p-k)}{E(p) E(p-k)} \right) \ .
	\label{equis}
\eeq  
The norm then can be written as
$Z = Z_A Z_{\psi} Z_{\phi}$, where $Z_A$ and $Z_\psi$ are constants
which cancel in all calculations and
\beq
Z_{\phi} = \int D\phi \, 
	\exp
	\left( - \frac{1}{4} \int dx \, dy \, 
	\partial\phi(x) \, S(x-y) \, \partial\phi(y) \right) \ .
\eeq 
The ``effective action'' for $\phi$ is given by
\beq
\label{18}
S(k) = G^{-1}(k) +\frac{e^2}{\pi} X(k) \ .
\eeq

We now compute the expectation value of the Hamiltonian in the
variational state.  Minimization of this quantity with respect to the 
variational parameters  will yield the ground state of the theory.  
In doing so one should be careful with properly regularizing the
ultraviolet singular operators that enter the Hamiltonian,

\beq
{\cal H}(x) = \frac{1}{2} E^2(x) + 
	i \bar{\psi}(x) \gamma_1 
	\left( \partial + i e A(x) \right) \psi(x) \ .
\eeq
The term to keep an eye on is the interaction term. In particular
the current $j(x) = \bar{\psi}(x) \gamma \psi(x)$
has to be regularized using the
point-splitting technique \cite{julian}.  
This point-splitting generates an additional term in the Hamiltonian
which is quadratic in $A$\cite{julian,Becher},
\beq 
i \bar{\psi} D_{\mu} \gamma^{\mu} \psi = 
	- \frac{\Lambda^2}{2 \pi} + 
	i : \bar{\psi} \partial_{\mu} \gamma^{\mu} \psi: - 
	e \, A :\bar{\psi} \gamma^1 \psi: - 
	\frac{e^2}{2 \pi} A^2 + \frac{ie}{2\pi}
	\partial A \ ,   
	\label{pepe}
\eeq 
where $\Lambda$ is the UV cutoff.  
The first term is a constant and it does not affect our
calculation. The last term is a total derivative and for this reason will also
be dropped.
We are therefore concerned with the 
calculation of the expectation of the regularized Hamiltonian,
\beq
\label{Hreg}
H = 	\int dx \left(\frac{1}{2} E^2(x) + 
	:i \bar{\psi}(x) \gamma_1 \partial \psi(x): + 
	e \, A(x) :\bar{\psi}(x) \gamma_1 \psi(x): + 
	\frac{e^2}{2\pi} A^2(x)\right) . 
\eeq
We proceed to calculate the expectation of (\ref{Hreg}) in the manner
outlined above for the calculation of the norm.  We find,
\beqa
\langle \frac{1}{2} \int dx \, E^2(x) \rangle 
	&=& 
	\frac{1}{4} \int \frac{dk}{2\pi} 
	\left[ G^{-1}(k) - G^{-2}(k) S^{-1}(k) \right] \ , \\
\langle i \int dx \, 
	:\bar{\psi}(x) \gamma_1 \partial \psi(x): \rangle
	&=&
	\lim_{\Lambda \to \infty}
	\left( 
	\int_{-\Lambda}^{\Lambda} 
	\frac{dk}{2\pi} 
	\frac{k^2}{\sqrt{k^2 + m^2(k)}}
	- \frac{\Lambda^2}{2 \pi} 
	\right) \ , \\
\langle e \int dx \, \bar{\psi}(x) \gamma A \psi(x) \rangle 
	&=& 
	\frac{e^2}{2 \pi}
	\int \frac{dk}{2 \pi} S^{-1}(k) \ , \\
\langle \frac{e^2}{2 \pi} \int dx \, A^2(x) \rangle 
	&=& 
	\frac{e^2}{4 \pi} \int 
	\frac{dk}{2 \pi} \left[ G(k) + S^{-1}(k) \right] .
\eeqa

Functional variation of the expectation of the Hamiltonian with
respect to the parameters $m(p)$ and $G^{-1}(p)$ yields two
minimization equations, 
\beqa
\frac{\delta \langle H \rangle}{\delta m(p)} &=& 
	\int \frac{dk}{8 \pi}
	\left( G^{-2}(k) + \frac{e^2}{\pi} \right)
	S^{-2}(k) \frac{\delta X(k)}{\delta m(p)}
	- \frac{p^2 m(p)}{\left(p^2 + m^2(p)\right)^{3/2}}
	= 0 \ , \\
\frac{\delta \langle H \rangle}{\delta G^{-1}(p)} 
	&=& 
	\frac{1}{4} S^{-2}(p)
	\left[
	\left(1 - \frac{e^2}{\pi} G^2(p)\right) X^2(p)
	- \frac{2 e^2}{\pi} X(p) G(p) 
	\right] = 0 \ .
\eeqa
Since ${\delta X / \delta m}\propto m$, the first equation 
is solved for $m(p)=0$. The second minimization equation
has the solution 
\beq
\left( \frac{e^2}{\pi} G(p) + X^{-1}(p) \right)^2 =
	\frac{e^2}{\pi} + X^{-2}(p) \ ,
\eeq 
where $X(p)$ is defined in (\ref{equis}).
With  $m(p)=0$ we have  $X(p) = 1/|p|$, 
and we can write the solution as,
\beq
\label{30}
\left( \frac{e^2}{\pi} G(p) + |p| \right)^2 = 
	\frac{e^2}{\pi} + p^2 \ . 
\eeq 
Summarizing, the result of our variational calculation is the vacuum 
wavefunctional (\ref{aa}) with the parameters 
\beq
\label{sol}
m(p) = 0 \ , 
	\;\;\;\;\;
	G(p) = \frac{\pi}{e^2} 
	\left(
	\sqrt{\frac{e^2}{\pi} + p^2} - |p|
	\right) .
\eeq
The calculations in this section rely upon the perturbative expansion
of the ``effective action'' for $\phi$ in (\ref{pepi}). To
improve on this we would have to resum this expansion.  
However, for $m(q)=0$ such a
re-summation is trivial. It can be shown explicitly that all
 diagrams higher than second order vanish and therefore our result is
exact at least in the sense that the solution we found is at least a
local minimum. We will see in the next section that the result is
indeed exact.

\section{Bosonized Formulation of the Schwinger Model.}

As is well known the Schwinger model is exactly solvable by the
bosonisation technique \cite{C75}.
In this section we will discuss the variational calculation in this
bosonized form.
We shall show that the solution obtained by the variational
method in the bosonized form is exact.
Further, we shall show that this
solution is equivalent to the wavefunctional obtained in the
previous section constructed in the fermionic Hilbert space.  

The bosonized form of the Hamiltonian density in the temporal
gauge ($A_0=0$) reads,
\beq
{\cal H}(x) = \frac{1}{2} E^2(x) + 
	\frac{1}{2} \left( p(x) - \frac{e}{\sqrt{\pi}} A(x) \right)^2
	+ \frac{1}{2} \, \left[ \partial \chi(x) \right]^2 \ ,
\eeq
which is supplemented by a Gauss' law constraint on physical states,
\beq
{\cal G}_B(x) = E(x) - \frac{e}{\sqrt{\pi}} \ \chi(x)=0 \ ,
\label{gb}
\eeq
where $E(x)$ and $A(x)=A_1(x)$ are the electric and photon fields as
defined in the previous section, $\chi(x)$ is the boson field, and
$p(x)$ its canonical momentum. 
Since the Hamiltonian is quadratic in terms of bosonic fields and
their conjugate momenta, the exact ground state wave functional can be
readily found,
\beq
\label{wfbexact}
\Psi_{0}[A,\chi] = 
	\exp\left[\frac{ie}{\sqrt{\pi}} 
	\int dx \, \chi(x) A(x) - 
	\frac{1}{2} \int dx \,  dy \, \chi(x) 	
	\Sigma^{-1}(x - y) \chi(y)\right],
\eeq
with $\Sigma^{-1}(k) = \sqrt{k^2 + e^2 / \pi}$ and the suffix $0$
indicates the exact ground state.

We now want to show that this result is recovered within the
variational calculation.
We base our variational Ansatz  upon a 
Gaussian,
\beq
\exp\left[ -\frac{1}{2} \int dx \, dy \, A(x) G^{-1}(x - y) A(y) - 
	\frac{1}{2} 
	\int dx \, dy \, \chi(x) \tilde{\Sigma}^{-1}(x - y) 
	\chi(y) \right] ,
\eeq
with $G(x-y)$ and $\tilde{\Sigma}$ as variational parameters.
Projecting this onto the gauge invariant subspace with the help of the
Gauss' law eq.(\ref{gb}) we have the gauge invariant trial state, 
\beqa
\label{wfb}
\Psi_B[A,\chi] &=& \int D\phi \ \exp\left[
	-\frac{1}{2} 
	\int dx \ dy \ A^{\phi}(x) G^{-1}(x - y) A^{\phi}(y) 
	\right. \non \\
	& & \left. 	
	- \frac{1}{2} \int dx \ dy \ \chi(x) 
	\tilde{\Sigma}^{-1}(x - y) \chi(y) + 
	\frac{ie}{\sqrt{\pi}} \int dx \ 
	\partial \phi(x) \chi(x) \right] \ .
\eeqa

Proceeding with the variational calculation in the usual way, we find
the expectation value of the Hamiltonian to be,
\beqa
\label{HB}
\langle H \rangle &=& 
	\frac{1}{4} \int \frac{dk}{2\pi} \left(G^{-1}(k) + 
	\frac{e^2}{\pi} \tilde{\Sigma}(k) \right)^{-1} 
	\non \\
	& & \times \,
	\left[ \frac{e^2}{\pi} G^{-1}(k)
	\tilde{\Sigma}(k) + \left(\tilde{\Sigma}(k) G(k)\right)^{-1} + 
	\frac{2 e^2}{\pi} + 
	k^2 \tilde{\Sigma}(k) G^{-1}(k) + \frac{e^4}{\pi^2} G(k)
	\tilde{\Sigma}(k) \right] 
	\nonumber \\
	&=& 
	\frac{1}{4} \int \frac{dk}{2\pi}
	\left(Y^2(k) + k^2 + \frac{e^2}{\pi}\right)
	Y^{-1}(k) \ ,
\eeqa
where
\beq
Y(k) = \tilde{\Sigma}^{-1}(k) + \frac{e^2}{\pi} G(k) \ .
	\label{uai}
\eeq
Variational minimization of (\ref{HB}) with respect to $Y(k)$ 
yields the required solution,
\beq
\label{ysol}
Y^2(k) = \left( \tilde{\Sigma}^{-1}(k) + 
	\frac{e^2}{\pi} G(k) \right)^2 =  
	\left( k^2 + \frac{e^2}{\pi} \right).
\eeq
After substituting into (\ref{wfb}) and integrating over $\phi$, 
we find that the best variational state is indeed (\ref{wfbexact}),
so the variational calculation gives the exact result. 

Note that our initial parameterization of the trial state turned out to
be somewhat redundant. The energy minimization determines only $Y$ but
not $G$ and $\tilde \Sigma$ separately. It is in fact true that
integrating over $\phi$ in equation (\ref{wfb}) we obtain a Gaussian
in $\chi$ with width $Y$. This redundancy however can be used to
establish the equivalence of this calculation with the calculation in
terms of the fermionic representation of the previous section.
The point is that we can choose $\tilde\Sigma$ to be the width of the
ground state in the free massless bosonic theory, 
\beq
\tilde\Sigma^{-1}=\sqrt{k^2} \ . 
\label{sigma}
\eeq
The exact ground state is then represented as a gauge projected
product of the vacuum wave functional of a theory of a massless boson
and a Gaussian functional of the gauge potential $A$. In fact for this
choice of $\tilde\Sigma$, eq.(\ref{ysol}) gives the same result for
the width of the gauge field Gaussian $G$ as the fermionic solution 
eq.(\ref{sol}). Also the free massless fermion by bosonisation is
equivalent to the free massless boson. Therefore the fermionic part of
the variational state of the previous section for $m=0$ is equivalent
to the free massless state for the bosonic field $\chi$.
We therefore see that our variational solution of the previous section
is equivalent to the solution obtained here, and is therefore an exact
ground state.

\section{Conclusions}
In this paper we have presented a variational analysis of the massless
Schwinger model. We have shown that there exists a simple and natural
way to extend the gauge invariant variational Ansatz of \cite{KK95} to
systems with fermions. We have also found that in this simple and well
understood system our variational Ansatz in fact includes the exact
ground state and therefore reproduces known exact results.

A noteworthy feature of the best variational state is that
the dynamical mass in the fermionic sector is not
generated,
so that in terms of fermionic part the state
seems to be trivial. In fact in 2 dimensional theory this is not
entirely surprising. 
If we were to consider a theory with more than one flavour, this would
have been an immediate corollary of the Coleman theorem. Since a
continuous (chiral) symmetry can not be broken in 2 D, the state
with nonvanishing dynamical mass should be strongly energetically
disfavoured.
In the one flavour theory which we are considering here, such an
a priori argument does not apply since the axial $U(1)$ symmetry is
anomalous. Nevertheless it is not unnatural that in terms of the
dynamical mass the ground states in one flavour and multiflavour
theories are similar.

On the other hand in the one flavour case the fermionic bilinear
condensate $\bar{\psi} \psi$ should be nonvanishing precisely due to
the same anomaly. It is therefore interesting to see whether our best
trial state leads to such a nonvanishing condensate.

It is simplest to calculate the condensate 
first in the bosonized version of the theory. 
The bosonization rules prescribe the identification \cite{C75},
\beq
\bar{\psi}(x) \psi(x) = - \frac{c \, e}{\sqrt{\pi}}
	\, {\cal N}_{e/\sqrt{\pi}} 
	\cos \left[ \sqrt{4 \pi} \chi(x) \right] \ , 
	\label{chita}
\eeq
where ${\cal N}_{e/\sqrt{\pi}}$ means normal ordering with respect to 
the free boson field of mass $e/\sqrt{\pi}$, and 
the prefactor is $c=\exp{(\gamma)}/2 \pi$, 
with $\gamma$ being Euler's
constant \cite{smilgaplb}. 
We need the average of (\ref{chita}) over the wavefunctional
(\ref{wfb}). After integrating over $A$ and $\chi$ we obtain
\beqa
\langle \bar{\psi}(x) \psi(x) \rangle
	&=&	
	- \frac{c \, e}{\sqrt{\pi}}
	\, {\cal N}_{e/\sqrt{\pi}} 
	\int D\phi \, \exp \left[  
	- \frac{1}{4} \int dy \, dz \, 
	\partial\phi(y) \, 
	\left(G^{-1}(y-z) + \frac{e^2}{\pi} \tilde{\Sigma}(y-z)\right) 
	\, \partial\phi(z)
	\right.
	\non \\
	&& \left.
	+ e \int dy \, \tilde{\Sigma}(x-y) \, \partial\phi(y)
	- \pi \, \tilde{\Sigma}(0)
	\right] \ . 
\label{cond1}
\eeqa
Using the definition (\ref{uai}), 
a further integration over $\phi$ yields
\beq
\langle \bar{\psi}(x) \psi(x) \rangle =
	- \frac{c \, e}{\sqrt{\pi}}
	\, {\cal N}_{e/\sqrt{\pi}} \,
	\exp \left[ - \pi Y^{-1}(x-x) \right]
\eeq
The exponential is UV divergent, but it cancels exactly with the
normal ordering factor, so we finally obtain 
\beq
\langle \bar{\psi}(x) \psi(x) \rangle =
	- \frac{e \, \exp{(\gamma)}}{2 \pi^{3/2}} \ , 
\label{cond2}
\eeq
which is the exact result \cite{all}.

The same calculation can be performed in the fermionic formalism.
We need to calculate
\beqa
\langle \bar{\psi}(x) \psi(x) \rangle
	&=&
	\int D\phi \, D\bar{\psi} \, D\psi \,
	\bar{\psi}(x) \psi(x) \, 
	\exp \left[
	- \frac{1}{4} \int dy \, dz \, 
	\partial\phi(y) G^{-1}(y-z) \partial\phi(z) 
	\right.
	\non \\
	&&
	\left.
	+ i \int dt \, dx \, 
	\bar{\psi}(x,t)	
	\gamma^{\mu} \left( i \partial_{\mu} - e a_\mu \right) 
	\psi(y,t) 
	\right] \ , 
\eeqa
with $a_0(x,t) = \phi(x) \delta(t)$ and $a_1(x,t)=0$.
Here we have used the path integral 
representation of the average over the
massless fermion vacuum state  eq.(\ref{pirulo}).
To perform the  integration over the fermions we can use the
results of \cite{sw}.
The fermionic integral is non-vanishing only for background fields
with instanton number $\pm 1$. Decomposing $a_\mu = \partial_\mu
\theta - \epsilon_{\mu \nu} \partial_\nu \varphi$, we obtain
\beqa
\langle \bar{\psi}(x) \psi(x) \rangle
	&=&
	\mbox{const.} \times
	\int D\phi \, 
	\exp \left[
	- \frac{1}{4} \int dy \, dz \, 
	\partial\phi(y) \, S(y-z) \, \partial\phi(z)
	\right.
	\non \\
	&&
	\left. 
	+ {1\over 2\pi}\int dy \, \log(x-y) \, \partial\phi(y)
	\right] \ . 
\eeqa
Here we have used 
\beq
\square \varphi(x,t) = - \partial \phi(x) \, \delta(t) 
	\Rightarrow
	\varphi(x,t=0) = 
	{1\over 2\pi}\int dy \log(x-y) \partial \phi(y) \ .
\eeq

With $\tilde\Sigma$ given by eq.(\ref{sigma}) this 
coincides with eq.(\ref{cond1}) and so the subsequent integration over
$\phi$ again reproduces eq.(\ref{cond2}).

We have thus seen that in this simple model the gauge invariant
variational Gaussian approximation works well. Extension of this
approach to four dimensional QCD with chiral symmetry breaking is 
the next challenging task.

\acknowledgements

The work of W.E.B. was supported in part by the U.S. Department of
Energy under grant DOE-91ER40651-TASKB.
The work of J.P.G. was supported by EC Grant
ARG/B7-3011/94/27. 
I.I.K. was partially supported by PPARC rolling grant PPA/G/O/1998/00567.
A.K. is supported by a PPARC advanced fellowship.

\end{document}